\documentclass[a4paper,UKenglish,cleveref, autoref, thm-restate]{lipics-v2021}
\hideLIPIcs
\usepackage{hyperref}
\usepackage{graphicx}
\usepackage{booktabs}
\usepackage{svg}
\usepackage{amsmath}
\usepackage{amssymb} 
\usepackage{float}
\usepackage{listings}
\usepackage{array}
\usepackage{amsmath}
\usepackage{tikz}
\usetikzlibrary{arrows.meta, positioning, shapes.geometric, calc}
\usepackage{lineno}

\usepackage{mathtools}

\title{ETH Flippers Approach to Parallel Reconfiguration of Triangulations: SAT formulation and Heuristics} %TODO Please add

\titlerunning{ETH Flippers Approach to Parallel Reconfiguration of Triangulations} %TODO optional, please use if title is longer than one line

\author{Lorenzo Battini}{ETH Zürich, Switzerland}{lbattini@student.ethz.ch}{https://orcid.org/0009-0009-7055-2277}{}%TODO mandatory, please use full name; only 1 author per \author macro; first two parameters are mandatory, other parameters can be empty. Please provide at least the name of the affiliation and the country. The full address is optional. Use additional curly braces to indicate the correct name splitting when the last name consists of multiple name parts.

\author{Marko Milenković}{ETH Zürich, Switzerland \and \url{https://sites.google.com/view/sprdalo} }{mmilenkovic@student.ethz.ch}{https://orcid.org/0009-0004-7925-7219}{}

\authorrunning{L. Battini and M. Milenković} %TODO mandatory. First: Use abbreviated first/middle names. Second (only in severe cases): Use first author plus 'et al.'

\Copyright{Lorenzo Battini and Marko Milenković} %TODO mandatory, please use full first names. LIPIcs license is "CC-BY";  http://creativecommons.org/licenses/by/3.0/

\ccsdesc[100]{Theory of Computation $\rightarrow$ Computational Geometry} %TODO mandatory: Please choose ACM 2012 classifications from https://dl.acm.org/ccs/ccs_flat.cfm 

\keywords{exact solution, heuristic, SAT solver, XOR clauses, computational geometry} %TODO mandatory; please add comma-separated list of keywords

\category{CG Challenge} %optional, e.g. invited paper

\relatedversion{} %optional, e.g. full version hosted on arXiv, HAL, or other respository/website
\supplement{https://github.com/Lbattini/ETH-flippers-CGSHOP2026}%optional, e.g. related research data, source code, ... hosted on a repository like zenodo, figshare, GitHub, ...
\acknowledgements{We would like to thank the Challenge organizers and other competitors for their time, feedback, and for making this whole event possible. This work was conducted as part of the Practical Work course at ETH Zürich under the supervision of Michael Hoffmann, whose guidance was invaluable. We sincerely thank Bernd Gärtner for providing travel support to attend the conference. We also acknowledge the ETH High Performance Computing group for providing the computational resources of the Euler cluster. }%optional

\nolinenumbers %uncomment to disable line numbering

\begin{document}

\maketitle

\begin{abstract}
We describe the algorithms used by the ETH Flippers team in the CG:SHOP 2026 Challenge. Each instance consists of a set of triangulations on a common point set, and the objective is to find a central triangulation that minimizes the total parallel flip distance to the input set. Our strategy combines an exact solver for small and medium-sized instances with a suite of heuristics for larger instances. For the exact approach, we formulate the problem as a SAT instance with XOR clauses to model edge transitions across multiple rounds, further optimized by lower bounds derived from exact pairwise distances. For larger instances, we use a greedy local search and edge-coloring techniques to identify maximal sets of independent flips. Our approach ranked second overall and first in the junior category, computing provably optimal solutions for 186 out of 250 instances.
\end{abstract}

\section{Introduction} \label{introduction}
Triangulations of point sets, in this context, maximal straight-line crossing-free graphs with endpoints in the point set, have many applications, for instance, in computer graphics and in reconstruction problems in geodesy.

A common problem is transforming a triangulation into another on the same point set using edge flips, replacing an internal edge whose incident triangles form a convex quadrilateral with its other diagonal. Hurtado, Noy, and Urrutia \cite{hurtado1998parallel} introduced parallel edge flips, which consist of flipping at the same time a set of edges such that none of the pairs of edges in the set shares a triangle.

The “CG:SHOP Challenge” (Computational Geometry: Solving Hard Optimization Problems) is a computational challenge competition about a hard geometric optimization problem. This year's challenge problem was proposed by Aichholzer, Dorfer, and Kramer. The goal is to find a central triangulation $C$ and, for each of the given $m$ triangulations $T_1, T_2, \ldots, T_m$ on the same pointset $P$, a sequence of parallel flips transforming $T_i$ into $C$, such that the sum of the lengths of these sequences is minimized. The solution methods are evaluated on a benchmark of 250 instances. We refer to \cite{survey} for further challenge details.

We present an exact solution method based on formulating the problem as a SAT problem with XOR clauses, which is used to solve the small and medium instances, as well as various heuristics that we used for the larger instances.

\begin{figure} [h]
\begin{center}
\includegraphics[scale=0.7]{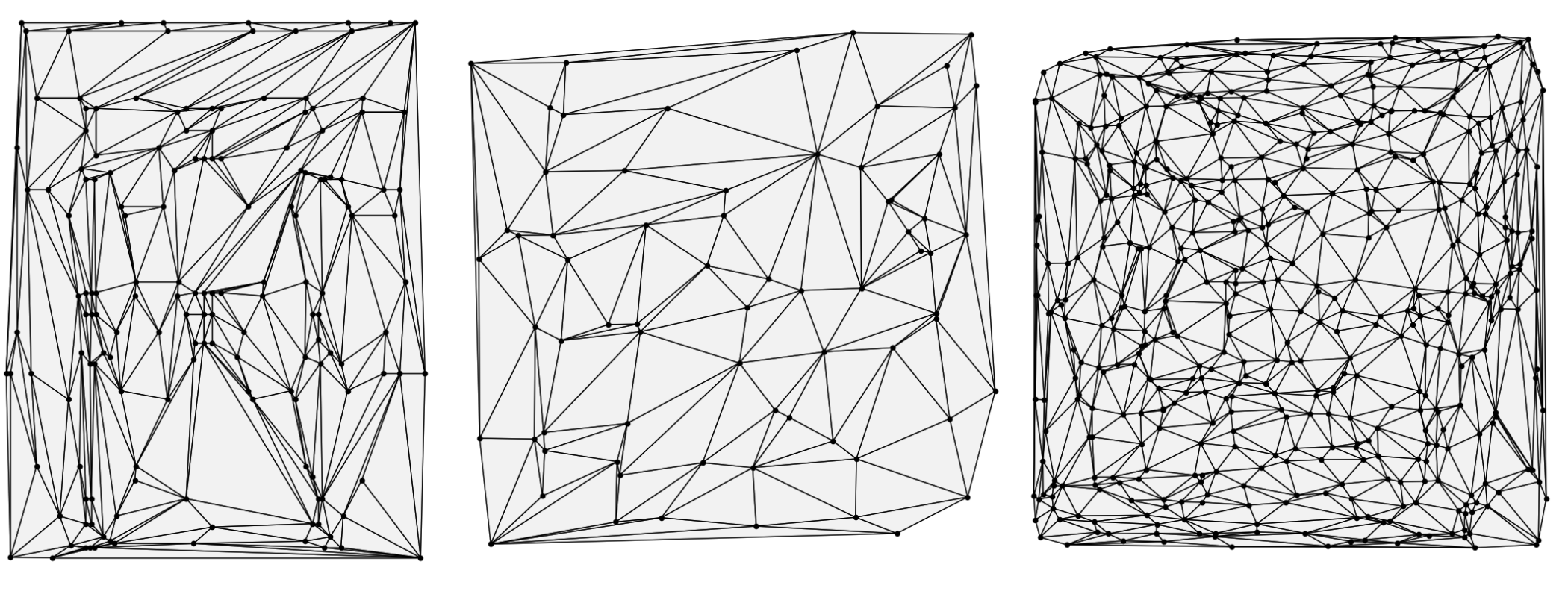}
\end{center}
\caption{Our best centers to instances \lstinline{random_instance_440_160_20}, \lstinline{woc-70-random-9a7d18d3}, \lstinline{rirs-500-50-23d00ec5}. }
\label{fig:vis}
\end{figure}

Our team, ETH flippers, ranked second overall and first in the junior category. We computed a provably optimal solution for 186 instances. Three further instances were solved, but a technical bug during our local validation caused us to stop at sub-optimal values. The largest instances we optimally solved featured twenty $320$-point triangulations, while the highest total parallel flip distance among our optimally solved instances is $92$. Notably, for this latter instance (\texttt{random\_instance\_440\_160\_20}), our solution was uniquely the best among all participating teams (see Figure \ref{fig:vis}).

\section{2-SAT formulation}\label{sec:2sat}

Given a set of $m$ triangulations, $\mathcal{T} = \{T_1, T_2, \dots, T_m\}$, we formulate a 2-SAT instance to search for a triangulation $C$ such that the distance from each $T_i \in \mathcal{T}$ to $C$ is at most one.

For every convex quadrilateral $q$ formed by a pair of facial triangles that share an edge in any $T_i$, we fix an arbitrary reference diagonal. Let the variable $x_q$ denote whether $C$ uses its reference diagonal and let the constant $b_q^{(i)}$ denote its state in $T_i$. To prevent invalid simultaneous flips, we add the clause $\neg(b_q^{(i)} \oplus x_q) \lor \neg(b_{q'}^{(i)} \oplus x_{q'})$ for all pairs of intersecting quadrilaterals $q, q'$ in each $T_i$. This determines the existence of a central triangulation $C$ in $\mathcal{O}(mn)$ time.

\section{SAT formulation with XOR clauses}

\subparagraph{XOR clauses.}

The most natural approach to modelling parallel flips is a round-by-round progression. We need to keep the state of each edge across intermediate parallel flips.

We model the state transition of an edge $e$ over $k$ parallel flips using XOR operations. Let $e^0, e^k \in \{0, 1\}$ denote the presence of $e$ in the initial and final triangulations, respectively. Let $\{q_1, \dots, q_g\}$ be the set of all potential convex empty quadrilaterals that share $e$ as a diagonal. We let $f_i^j \in \{0, 1\}$ indicate whether a flip is performed on $q_i$ during step $j$.

Because the final state of edge $e$ after $k$ parallel flips is simply its initial state XORed with all intermediate flips across its potential quadrilaterals $q_{1 \dots g}$, we add the constraint \begin{equation*}e^0 \oplus e^k \oplus \bigoplus_{j=1}^k \bigoplus_{i=1}^g f_i^j = 0\end{equation*}
We decided to use the CryptoMiniSat5 \cite{cryptominisat} solver, since it supports inputs with XOR clauses.

\subsection{Formulation}
We extend the previous formulation to search for a center $C$ whose distance to each $T_i \in \mathcal{T}$ is upper bounded by a given target $d[i]$.

\subparagraph{Geometric Preconditions for Flips.}

Let $\mathcal{Q}$ be the set of all convex empty quadrilaterals defined on the pointset, which we identified via spatial hashing \cite{Teschner2003}. Let the variable $e_{uv}^t$ denote the presence of edge $(u,v)$ at step $t$. For a  flip $f_{q}^t\in \{0,1\}$ to be geometrically valid at step $t$, the four boundary edges $\partial q$ of its corresponding quadrilateral $q \in \mathcal{Q}$ must exist in the prior step. We enforce this constraint with implication $f_{q}^{t} \implies \bigwedge_{(u,v) \in \partial q} e_{uv}^{t-1}$.

\subparagraph{Target Equivalence with Center C.}

We introduce global boolean variables $c_{uv}$ representing the edges of the center $C$. We enforce consistency at the final step $k$ via the equivalence $e_{uv}^{k} \iff c_{uv}$. This link forces the final state of every flip sequence to be identical to the global center, effectively merging the independent SAT sub-problems to a shared solution.
\subparagraph{Edge Transition Constraints.}

We must ensure that the triangulation changes only through valid flip operations. An edge $(u,v)$ changes its status (present or absent) between parallel flips $t-1$ and $t$ if and only if a flip is performed on a quadrilateral where $(u,v)$ serves as the diagonal.

Let $\mathcal{Q}_{uv}$ be the set of all convex empty quadrilaterals for which $(u,v)$ is a diagonal. The relationship between the edge's state at consecutive parallel flips and the flip variables is enforced via an XOR constraint:
$  e_{uv}^{t-1} \oplus e_{uv}^{t} \oplus \bigoplus_{q \in \mathcal{Q}_{uv}} f_{q}^{t} = 0$.

For edges on the convex hull, the set $\mathcal{Q}_{uv}$ is empty, effectively enforcing that these edges remain constant throughout the transformation.

\subparagraph{Constraints on Simultaneous Flips.}

We must enforce strict independence between simultaneous operations. Specifically, for any subset of four vertices $\{v_1, v_2, v_3, v_4\}$, there can be at most one active flip operation at time $t$ involving three of these vertices. This prevents structural conflicts where multiple flips compete for the same triangle.

We iterate through all quadruplets of vertices and identify the set of potential flip variables $\mathcal{F}_{v_{1..4}}^{t}$ of size $r$ associated with them. We then enforce an \textit{At-Most-One} (AMO) constraint on this set $\sum_{f \in \mathcal{F}_{v_{1..4}}^{t}} f \le 1$.
There are multiple ways to model this constraint using a linear number of clauses \cite{Sinz2005}. The ladder encoding that we implemented uses $2r-1$ variables and $3r-2$ clauses, compared to $r$ variables and $\binom{r}{2}$ clauses. Due to the $\mathcal{O}(n^4)$ possible number of convex empty quadrilaterals, clauses generation becomes a computational bottleneck.

\subparagraph{Initial State and Boundary Conditions. }

We initialize the SAT model for each input triangulation by fixing $e_{uv}^{0}$ to true if the edge $(u,v)$ exists in it and false otherwise. Furthermore, since parallel flips only affect internal diagonals of convex quadrilaterals, the convex hull edges $\mathcal{H}$ remain invariant. We hard-code $e_{uv}^{t}$ to true for all $(u,v) \in \mathcal{H}$ across all steps $t \in \{1, \dots, k\}$, effectively pruning the search space by preventing flips of boundary edges.

\subparagraph{Pruning clauses. } \label{pruningWithMight}
To significantly prune the search space, we precompute the minimum flip distance required to introduce each edge into triangulation $T_i$ using an iterative breadth-first reachability search. We then omit all SAT variables and clauses associated with an edge for any time step prior to this theoretical minimum.

\subsection{Finding convex empty quadrilaterals} \label{findingConvexEmptyQuadr}

Before constructing the clauses, a necessary preliminary step is the identification of all empty convex quadrilaterals, as these form the basis of our formulation. The problem of finding such quadrilaterals is fundamental in computational geometry and has been extensively researched \cite{dobkin1990finding}.

It is important to note that, theoretically, the number of empty convex quadrilaterals can reach $\Theta(n^4)$. A classic example of this worst-case scenario is a set of points in convex position (e.g., vertices of a regular polygon). In such instances, the high volume of quadrilaterals makes a certain degree of computational expense unavoidable. However, in practice, this behavior is rarely observed in randomly generated datasets. Furthermore, if the points do lie in a convex position, the underlying problem actually simplifies, as this specific case has been thoroughly researched in a single flip setup \cite{Li2023}.

\subparagraph{K-d trees}

A standard algorithmic approach for optimizing range queries and emptiness checks in computational geometry is the $k$-d tree \cite{bentley1975multidimensional} (k-dimensional tree). This binary space partitioning data structure recursively subdivides the Euclidean plane into disjoint rectangular regions using axis-aligned hyperplanes.

To use a k-d tree for detecting empty convex quadrilaterals, the algorithm proceeds as follows:
\begin{enumerate}
    \item \textbf{Construction:} The tree is built by recursively splitting the point set $P$ along alternating axes (x and y) at the median coordinate. This creates a balanced structure where each node represents a specific rectangular region of the plane and the points contained within it.
    \item \textbf{Emptiness Query:} To check if a candidate quadrilateral $q$ contains any interior points, we perform a range search. Instead of checking every point in $P$, the traversal relies on the Axis-Aligned Bounding Box (AABB) of $q$.
    \item \textbf{Pruning:} As the algorithm traverses the tree, it checks for intersections between the node's region and $q$'s bounding box:
    \begin{itemize}
        \item If a node's region is strictly disjoint from $q$'s bounding box, the entire subtree is pruned (discarded).
        \item If a node's region is fully contained within $q$, the search terminates immediately as the emptiness condition is violated.
        \item If the region partially intersects $q$, the search proceeds recursively to the child nodes.
    \end{itemize}
\end{enumerate}

\subparagraph{Spatial Hashing}

While the k-d tree offers a robust worst-case query time of $\mathcal{O}(\sqrt{N})$\cite{deberg2008computational} for orthogonal range searches and adapts well to highly clustered or skewed data distributions, we intentionally opted for the Uniform Grid Partitioning (Spatial Hashing \cite{teschner2003optimized}) approach for this project \cite{bentley1979data}.\\
First, the k-d tree introduces significant overhead due to pointer-based traversal and recursive branching logic. Second, and most importantly, our dataset exhibits a quasi-uniform distribution. In such a scenario, the adaptive balancing of a k-d tree is redundant. The Uniform Grid leverages this uniformity to provide an expected query time of $\mathcal{O}(1)$ by mapping spatial coordinates directly to array indices. This avoids the $\log N$ traversal depth of the tree and benefits from superior CPU cache locality, making it the more efficient choice for high-frequency emptiness checks in this specific competition environment.\\
We decompose the axis-aligned bounding box of the entire instance into a regular grid of cells. During the preprocessing phase, each point is bucketed into its corresponding cell based on its coordinates. Consequently, when verifying whether a candidate shape (e.g., a triangle or quadrilateral) is empty, we calculate the shape's bounding box and identify the specific grid cells it intersects. The search for interior points is then restricted exclusively to these relevant cells. This approach drastically prunes the search space, reducing the average-case complexity of the inclusion query significantly compared to a naive linear scan.\\
Regarding the computational complexity, the construction of the grid structure incurs a one-time preprocessing cost of $\mathcal{O}(N)$, as it requires a single linear pass to assign each of the $N$ points to its respective cell.\\
For the query phase (checking if a shape is empty), the worst-case time complexity remains $\mathcal{O}(N)$ in pathological scenarios where points are highly clustered into a single cell. However, in many instances of the dataset the points are distributed quite uniformly across the grid. If the grid dimensions are chosen such that the number of cells is proportional to $N$, the expected number of points per cell is constant ($\mathcal{O}(1)$). Consequently, the expected time complexity for an emptiness query is reduced to $\mathcal{O}(1)$, or more precisely, it becomes proportional to the area of the query shape relative to the grid density, rather than the total number of points.\\
To balance things, the dimensions of the grid cells are determined dynamically based on the dataset density. Specifically, the side length $s$ of each square cell is calculated as:
\begin{equation}
    s = \sqrt{\frac{(x_{max} - x_{min}) \cdot (y_{max} - y_{min})}{N}}
\end{equation}
where $x_{max}, x_{min}, y_{max}, y_{min}$ define the global bounding box of the point set, and $N$ represents the total number of points.\\
This parametrization implies that the total number of grid cells covers the same area as the bounding box divided by $N$. Under the assumption of a uniform distribution, this results in an expected density of approximately one point per cell. This heuristic efficiently balances the trade-off between the grid traversal overhead (too many empty cells) and the collision resolution cost (too many points per cell).

\section{SAT formulations without XOR clauses} \label{satWithoutXor}
\begin{figure} [H]
\begin{center}
\includegraphics[scale=1]{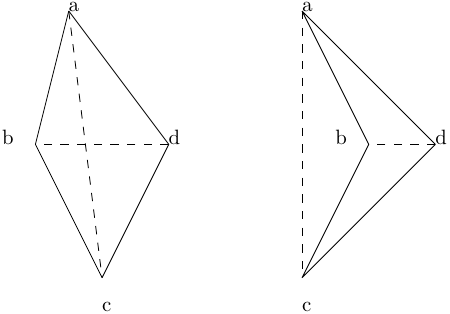}
\end{center}
\caption{A convex quadrilateral to the left, and a non convex quadrilateral to the right}
\label{exampleQuadr}
\end{figure}

\subparagraph{Formulation}
% We extend the previous formulation to arbitrary distances to the center $C$.
% More precisely, given a set of $m$ triangulations, $\mathcal{T} = \{T_1, T_2, \dots, T_m\}$, our objective is to find a single central triangulation $C$ such that the distance from $T_i \in \mathcal{T}$ to $C$ is at most $d_{center}[i]$.
We present an alternative formulation that does not use XOR clauses.\\
By step $s$ we refer to the $s$-th parallel flip, starting from $s=0$. 
% In this formulation, we want to decide if it is possible to flip each of the triangulations $T_0,...,T_m$ to a common center, with $d_{center}[i]$ steps for $T_i$ each.\\
$n$ is the number of points, $m$ is the number of triangulations.\\
We use the following boolean variables:
\begin{itemize}
\item $\prescript{}{s}{p_e^i}$ to denote whether edge $e=(u,v)$ is present at step $s$ in triangulation $T_i$. We also equivalently use $\prescript{}{s}{p_{uv}^i}$
\item $\prescript{}{s}{f_e^i}$ to denote whether edge $e=(u,v)$ is flipped at step $s$ in triangulation $T_i$. We also equivalently use $\prescript{}{s}{f_{uv}^i}$
\end{itemize}
To ensure that, after all flips, all input triangulations coincide, for the last step $d_{center}[i]$ we use the same variables $\prescript{}{d_{center}[0]}{p_{ab}^0}$ for all input triangulations. The choice of using $0$ as the triangulation index in those variables is arbitrary.\\
To reduce the number of clauses and variables, not all of these variables are present, but only those that are used in one of the clauses described below.\\
To describe the clauses used, we refer to an example quadrilateral with boundary edges $ab,bc,cd,da$ and diagonals $ac$ and $bd$, see \autoref{exampleQuadr} for an example.\\
We use the array $might$ defined in \autoref{pruningWithMight}.\\
% The array \texttt{mightBePresentFromStep} stores at entry \texttt{mightBePresentFromStep[i][e]} the first step at which edge $e$ might be present in triangulation $T_t$. We describe how it is computed in \autoref{computingMightBePresent}.\\
We say that a quadrilateral with boundary edges $B=\{ ab,bc,cd,da \}$ might be present at step $s$ if $\forall e \in B \; might[e][i] \leq s$.\\ 
At step $0$, edge $e$ is flippable if and only if it is the diagonal of an empty convex quadrilateral.\\
At step $s>0$, edge $e$ cannot be flipped if there is no empty convex quadrilateral with $e$ as a diagonal which might be present at step $s$.\\
We say that edge $e$ might be in the center if $\forall i \; might[e][i] \leq d_{center}[i]$.\\
The array $minDistanceFromCenter$ stores for each $e$ the parallel flip distance from $e$ to the closest edge which might be in the center.
We describe how it is computed in \autoref{computingMinDistFromCenter}.\\
The clauses used (in CNF) are:
\begin{itemize} 
\item $s \in [0,d_{center}[i]]$ \\
We must have a valid triangulation at all steps. This means that in each triangulation, at every step, there can be no pair of crossing edges $e_0,e_1$. \\
From this we get the clause $(\overline{\prescript{}{s}{p_{e_0}^i}} \vee \overline{\prescript{}{s}{p_{e_1}^i}})$ for all edges $e_0,e_1$ that cross each other, might be present at step $s$ and are not too far from the center. More precisely, it is added at steps $[\max(might[e_0][i],might[e_1][i]), \\ \min(maxS,d_{center}[i]-\max(minDistanceFromCenter[e0],minDistanceFromCenter[e1])]$. $maxS=d_{center}[i]$ if $i=0$, and $maxS=d_{center}[i]-1$ otherwise, since index $0$ was arbitrarily chosen as the index for the center variables.\\
We have at most $1$ such clause for each of the $\binom{n}{4}
$ quadruplet of points, for each triangulation, for each step $s>0$. A bound on the number of such clauses is therefore $\mathcal{O}(\sum_{i=0}^m (d_{center}[i]-1) n^4)$.
%every pair of edges $e_0,e_1$ that cross each other (in the same triangulation and step), 
\item $s=0$
\begin{enumerate}
\item For each flippable diagonal $e=\{a,c\}$ in the input triangulation $T_i$, either we flip it and we have the other diagonal at the next step, or we do not flip it and the edge $e$ is present at the next step. In CNF, we get
$ (\prescript{}{0}{f_{ac}^i} \vee \prescript{}{1}{p_{ac}^i}) \wedge
(\overline{\prescript{}{0}{f_{ac}^i}} \vee \prescript{}{1}{p_{bd}^i}) \wedge (\overline{\prescript{}{1}{p_{ac}^i}} \vee \overline{\prescript{}{1}{p_{bd}^i}})$.\\
%TODO in principle the last clause in this block (not present not present) is already present in the non crossing clauses, but it's duplicated in the code too, so I'll leave it
We have $\Theta(mn)$ such clauses.
%\item If \texttt{mightBePresentFromStep[t][idx(ac)]>1}, we have the clause $\overline{\prescript{}{1}{p_{ac}^i}}$
%\item If \texttt{!isFlippable[idx(ac)]}, meaning that edge $ac$ is in no convex quadrilateral at step $0$, we have the clause $\overline{\prescript{}{0}{f_{ac}^i}}$
% \texttt{!isFlippable[idx(ac)]} 
\item If $\{a,c\}$ is not flippable and $might[a,c][i]=0$, we have the clause $\prescript{}{1}{p_{ac}^i}$.\\
We have $\mathcal{O}(mn)$ such clauses.
\item For all pairs of edges $e$ and $e'$ that share a triangle in $T_i$ and might be flipped at step $0$, $\prescript{}{1}{\overline{f_e^i}} \vee \prescript{}{1}{\overline{f_{e'}^i}}$. \\
We have $\Theta(mn)$ such clauses.
\end{enumerate}
\item $s \in [1,d_{center}[i]-1]$
\begin{enumerate}
\item For each edge $e$ that might be flipped at step $s$, if $might[e,i]\leq s$, we have the clause $(\overline{\prescript{}{s}{p_{e}^i}} \vee \prescript{}{s}{f_{e}^i} \vee \prescript{}{s+1}{p_{e}^i})$. This ensures that if $e$ is present at step $s$ and it is not flipped at step $s$, it will be present at step $s+1$. If $might[e][i] \leq s$, but $e$ cannot be flipped at step $s$, we instead add the simplified clause $(\overline{\prescript{}{s}{p_{e}^i}} \vee \prescript{}{s+1}{p_{e}^i})$.\\
We have at most $1$ such clause for each edge in the point set, for each triangulation, for each step $s>0$. A bound on the number of such clauses is therefore $\mathcal{O}(\sum_{i=0}^m (d_{center}[i]-1) n^2)$, since we have at most $\binom{n}{2}
$ pairs of points in the point set. 
\item For each edge $e$ that might be flipped at step $s$, we have the clause $({\prescript{}{s}{p_{e}^i}} \vee \overline{\prescript{}{s}{f_{e}^i}} )$. This ensures that we cannot flip edges which are not present.\\
We have at most $1$ such clause for each edge in the point set, for each triangulation, for each step $s>0$. A bound on the number of such clauses is therefore $\mathcal{O}(\sum_{i=0}^m (d_{center}[i]-1) n^2)$.
%\item For each edge $e$ that belongs to no convex quadrilateral which might be present at step $s$, we add the clause $\overline{\prescript{}{s}{f_{e}^i}}$
\item For each of the input triangulations $T_i$, for each convex empty quadrilateral $q$ in the point set, and each of the two diagonals $e$ in it, if all the edges in $q$ and the diagonal $e$ are present, and $e$ is flipped, then the other diagonal $bd$ has to be present at step $s+1$.  We do not add this clause if at least one of the edges in $q$ or $e$ cannot be present at step $s$. If $minDistanceFromCenter[e_1] \geq d_{center}[i]-s$ we instead add clause \ref{cantFlipClause}, since flipping the edge would not lead to a valid assignment. In CNF, we get:
%either some edges (of $q$ or $e$) are missing, or the same constraints described at step 0) hold. In CNF, we get
%$$ (\overline{\prescript{}{s}{p_{ab}^i}} \vee \overline{\prescript{}{s}{p_{bc}^i}} \vee \overline{\prescript{}{s}{p_{cd}^i}} \vee \overline{\prescript{}{s}{p_{da}^i}} \vee \overline{\prescript{}{s}{p_{ac}^i}} \vee \prescript{}{s}{f_{ac}^i} \vee \prescript{}{s+1}{p_{ac}^i}) \wedge
$$
(\overline{\prescript{}{s}{p_{ab}^i}} \vee \overline{\prescript{}{s}{p_{bc}^i}} \vee \overline{\prescript{}{s}{p_{cd}^i}} \vee \overline{\prescript{}{s}{p_{da}^i}} \vee \overline{\prescript{}{s}{p_{ac}^i}} \vee \overline{\prescript{}{s}{f_{ac}^i}} \vee \prescript{}{s+1}{p_{bd}^i}) 
$$
We have at most $2$ such clauses for each empty convex quadrilateral in the point set, for each triangulation, for each step $s>0$. A bound on the number of such clauses is therefore $\mathcal{O}(\sum_{i=0}^m (d_{center}[i]-1) n^4)$, since we have at most $\binom{n}{4}
$ quadrilaterals in the point set. 
% few examples I tested, for instance they are $28869$ for an example instance with $100$ points. Moreover, for $d_{center}$ small we can reduce the number of such clauses by only considering edges that are at $d_{center}$ distance from some of the initial edges (at step $2$, for instance, there are only $\mathcal{O}(n)$ potential edges in a fixed triangulation).\\
% We also add a $\prescript{}{s+1}{p_{uv}} \vee \overline{\prescript{}{s}{p_{uv}}}$ clause for each edge $uv$ that is not present in any empty convex quadrilateral.
\item\label{cantFlipClause} For each of the input triangulations $T_i$, for each empty non convex quadrilateral $q$ in the point set that contains a single diagonal $e$ (since it is not convex, the other diagonal is outside the quadrilateral), if all the edges in $q$ are present, $e$ cannot be flipped. We don't add this clause if at least one of the edges in $q$ or $e$ cannot be present at step $s$. In CNF, we get:
$$
(\overline{\prescript{}{s}{p_{ab}^i}} \vee \overline{\prescript{}{s}{p_{bc}^i}} \vee \overline{\prescript{}{s}{p_{cd}^i}} \vee \overline{\prescript{}{s}{p_{da}^i}} \vee \overline{\prescript{}{s}{f_{ac}^i}}) 
$$
We have at most $1$ such clause for each empty non convex quadrilateral in the point set, for each triangulation, for each step $s>0$. A bound on the number of such clauses is therefore $\mathcal{O}(\sum_{i=0}^m (d_{center}[i]-1) n^4)$, since we have at most $\binom{n}{4}
$ quadrilaterals in the point set. 
\item For each of the input triangulations $T_i$, for each pair of flippable edges in the point set $e=\{u,v\}$ and $e'=\{u,y\}$ that share a vertex $u$, either the edge connecting the other vertices is missing, or they cannot both be flipped at step $s$. In CNF, we get $\overline{\prescript{}{s}{p_{vy}^{i}}} \vee \prescript{}{s}{\overline{f_e^i}} \vee \prescript{}{s}{\overline{f_{e'}^i}}$.\\
Since for each of the $\mathcal{O}(n^2)$ edges in the point set, there are $\mathcal{O}(n)$ edges that share a vertex with it, a bound on the number of such clauses is $\mathcal{O}(\sum_{i=0}^m (d_{center}[i]-1) n^3)$.
\item For each edge $e$ such that $minDistanceFromCenter[e]>d_{center}[i]-s$, we add the clause $\overline{\prescript{}{s}{p_e^t}}$.\\
We have at most $1$ such clause for each edge in the point set, for each triangulation, for each step $s>0$. A bound on the number of such clauses is therefore $\mathcal{O}(\sum_{i=0}^m (d_{center}[i]-1) n^2)$.
% \item To ensure that, after all flips, all input triangulations coincide, for the last step $d_{center}[i]$ we use the same variables $\prescript{}{d_{center}[0]}{p_{ab}}$ for all the input triangulations.
\end{enumerate}

\end{itemize}
% \subsection{Computing \texttt{mightBePresentFromStep}} \label{computingMightBePresent}
% We first initialize the array by setting \texttt{mightBePresentFromStep[i][e]=0} for all the edges that are initially present in $T_t$ and \texttt{mightBePresentFromStep[i][e]=1} for all the edges which correspond to the other diagonal of an empty convex quadrilateral which is initially present in $T_t$.\\
% Then we iteratively fill in the other values. To do so, for all steps $s \in [1,d_{center}[t]-1]$, for each edge $e_0$ such that \texttt{mightBePresentFromStep[i][$e_0$]$\leq s$}, for each $e_1$ which corresponds to the other diagonal of $e_0$ in an empty convex quadrilateral, such that \texttt{mightBePresentFromStep[t][$e_1$]} hasn't been assigned a value yet, we check if all the boundary edges of the empty convex quadrilateral might be present at step $s$, and, if so, set  \texttt{mightBePresentFromStep[i][$e_1$]=s+1}.

\subparagraph{Computing minDistanceFromCenter} \label{computingMinDistFromCenter}
% We first initialize the array by setting $minDistanceFromCenter[e]=0$ for all the edges $e$ that might be in the center.\\
% Then we iteratively fill in the other values. To do so, for all steps $s \in [0,(\max_i d_{center}[i])-1]$, for each edge $e_0$ such that $minDistanceFromCenter[e_0]\leq s$, for each $e_1$ which corresponds to the other diagonal of $e_0$ in an empty convex quadrilateral, such that $minDistanceFromCenter[e_1]$ hasn't been assigned a value yet, we check if all the boundary edges of the empty convex quadrilateral have minimum distance from center $\leq s$, and, if so, set  $minDistanceFromCenter[e_1]=s+1$.
%TODO Explain the pruning for edges that are not present at some step here rather than in the previous subsection; explain here 

We initialize $minDistanceFromCenter[e] = 0$ for all edges $e$ that might be in the center, and $minDinstanceFromCenter[e] = \infty$ otherwise. The values are updated iteratively: for each step $s \in [0,(\max_i d_{center}[i])-1]$, we examine every potential convex quadrilateral formed by vertices $\{u, x, v, y\}$ with an existing diagonal $(u,v)$ such that $minDistanceFromCenter[u,v] \le s$. If the boundary edges $B = \{(u,x), (x,v), (v,y), (y,u)\}$ satisfy the condition $\forall e \in B, minDistanceFromCenter[e] \le s$, it implies the quadrilateral is formable at step $s$. Consequently, the dual diagonal $(x,y)$ becomes reachable in the next step, and we update $minDistanceFromCenter[x,y] = \min(minDistanceFromCenter[x,y], s+1)$.

\subparagraph{SAT solvers}
Since this formulation does not use XOR clauses, any SAT solver can be used. 
With this formulation, we mainly used Kissat, see the paper by Armin Biere, Tobias Faller, Mathias Fleury, Nils Froleyks, and Florian Pollitt \cite{BiereFallerFazekasFleuryFroleyksPollitt-SAT-Competition-2024-solvers}. Since Kissat by itself is a sequential solver, we used the parallelism on the Euler cluster to either solve multiple instances in parallel, or different sequences of distances in parallel. We also used CryptoMiniSat5 \cite{cryptominisat} since it is the only parallel solver we were able to run on Euler.\\
We also attempted to use Mallob, a distributed platform for automated reasoning in modern large-scale HPC and cloud environments, see the paper by  Dominik Schreiber and Peter Sanders \cite{schreiber2024mallobsat}. It allows to run in parallel Kissat and other SAT solvers. However, we did not manage to make it run on Euler, while locally it runs sequentially on larger instances because of the limited amount of RAM available.
% Locally we also used mallob, but unfortunately we did not manage to make it run on euler, so ultimately it was not very useful (locally RAM limitations mean that on larger instances it essentially runs sequentially).

\subsection{Finding empty quadrilaterals} \label{findingEmptyQuadr}
The formulation described in \autoref{satWithoutXor} also needs non convex empty quadrilaterals. Although in principle the method described in the previous section could have been used to solve this problem, we did not find the time to do so. Here we present the slower method that we used.\\
First, we compute for each triplet of points $a,b,c$ whether it is an empty triangle or not by checking if any point is strictly inside the triangle $abc$. This requires $\mathcal{O}(n^4)$ time since there are ${\binom{n}{3}} = O(n^3)$ triplets of points, and for each of them we check if any of the $n$ points is strictly inside.\\
We also compute which pairs of points $a,b$ contain no point within the segment $ab$. \\
%TODO maybe put a label here and reference this figure
Then for each of the $\binom{n}{4}$ quadruplets of points, we sort them in counter clockwise order $a,b,c,d$. We then check if the two diagonals $ac$ and $bd$ are inside the quadrilateral. If $ac$ is outside, the quadrilateral is made up of triangles $t_0=\{b,d,a \}$ and $t_1=\{b,d,c \}$. Otherwise, the quadrilateral is made up of triangles $t_0=\{a,c,b \}$ and $t_1=\{a,c,d \}$. We store the triangles as array, with points ordered as we just defined them.\\
Then the quadrilateral is empty if and only if $t_0$, $t_1$ and the segment $t_0[0] t_0[1]$ are empty.\\
The quadrilateral is convex if both diagonals are inside and not convex otherwise.\\
Overall, the time required is $\Theta(n^4)$.\\
The computed empty triangles and empty quadrilaterals are stored in a file so that they only have to be computed once per instance.

\section{Exact solving}
\subparagraph{Lower bound.}
To establish a lower bound, we first compute exact pairwise parallel flip distances $D_{i,j}$ using our SAT formulation. Any valid central triangulation at distances $d_k$ to inputs $T_k$ must satisfy the triangle inequality $d_i + d_j \ge D_{i,j}$ for all pairs $i,j$. To test if a target total distance $S$ is feasible, we find all integer assignments $(d_1, \dots, d_m)$ summing to $S$ that satisfy these pairwise constraints via recursive backtracking. For instances with $m=20$ triangulations, we use some further pruning techniques. We compute the minimum parallel flip distance $s_{dist}(i)$ for the last $i$ triangulations, for $i \geq i_{start}$, where we used $i_{start}=11$ for most instances since it was experimentally found to be sufficiently fast to compute. Then we use those suffix distances by pruning a branch whenever $S-\sum_{k=1}^{i} d_k < s_{dist}(i)$.
Moreover, after the first $10$ and $15$ distances have been assigned recursively, we prune the branch if the prefix is not satisfiable.

\subparagraph{Choosing the distribution of distances. }

The upper bound search is existential. Consequently, prioritizing candidates that are  more likely to be satisfiable can significantly accelerate the search.

Let $d = (d_1, \dots, d_n)$ represent a potential assignment vector. We define the scoring function $\mathcal{S}(d)$ as the sum of absolute differences between all pairs of elements: $\mathcal{S}(d) = \sum_{i=1}^{n-1} \sum_{j=i+1}^{n} |d_i - d_j|$. The candidate distributions are processed in ascending order of $\mathcal{S}(d)$. 

Empirical analysis reveals that this heuristic is remarkably effective for randomly generated instances. In the majority of cases, the satisfiable distribution corresponds to the candidate with the minimal $\mathcal{S}(d)$ score. Even in cases where the global minimum is not the solution, the valid assignment is consistently found among the top-ranked candidates.

\begin{conjecture}
    For a randomly generated triangulation instance under the standard uniform distribution model, ordering of candidate distance vectors by increasing $\mathcal{S}(x)$ minimizes the expected number of queries required to find a satisfiable assignment.
\end{conjecture}

\section{Heuristics}
\subparagraph{Transforming a triangulation into another.}
To transform a triangulation $T$ into another $T'$, we flip edges such that the number of intersecting edges decreases. We now describe how, at each step, we choose a subset of independent edges to flip among the ones that reduce the number of intersections.

Firstly, we choose edges that, if flipped, introduce an edge that is present in $T'$. To do so, we start by creating a graph containing those edges, where the nodes correspond to faces incident to them. Then we compute an edge coloring of this graph using the Misra and Gries algorithm \cite{MISRA1992131} from the Boost Graph Library \cite{edgeColoringBGL}. Finally, we flip the edges corresponding to the color class of maximum cardinality, and all the isolated edges. Since the boundary edges cannot be flipped, and the bounded faces are triangles, the maximum degree $\Delta$ is at most $3$. Since the Misra and Gries algorithm uses at most $\Delta+1 \leq 4$ colors, at least one fourth of the edges in the graph are flipped.

Secondly, we greedily pick the edges that are independent from all the edges that have already been picked, in order of non-increasing reduction of the number of intersections.

This transforms a triangulation into the other in a finite number of steps, since it can be seen as an adaptation to parallel flips of the algorithm described by Hanke, Ottmann, and Schuierer \cite{hanke1996theEdge}. Given the flip sequence $\mathcal{F}=\{F_1,F_2,...,F_{|\mathcal{F}|-1}\}$ that we obtain with the described heuristic, let $\{\tau_0=T,\tau_1,...,\tau_{|\mathcal{F}|-1}=T'\}$ be the corresponding sequence of triangulations, where $\tau_{k+1}$ is the triangulation we obtain after the parallel flip $F_k$. The reversed flip sequence $\mathcal{F'}$ transforms $T'$ into $T$ and it can be obtained by reversing the order of the parallel flips, and replacing each individual flip $e \rightarrow e'$ with $e' \rightarrow e$.

We can obtain other heuristic flip sequences $\mathcal{F}_k$, with $k \in [0,|\mathcal{F}|-1]$, by concatenating the heuristic flip sequence that turns $T$ into $\tau_k$ with the reversed flip sequence that turns $T'$ into $\tau_k$. 

Since these sequences are heuristic, their length might differ, and so we try multiple values of $k$ and choose the shortest flip sequence.

We empirically observed that trying all intermediate $k$ values typically reduces each triangulation's distance by 1. Therefore, to maintain computational efficiency on large instances, we only do this when the total distance is within $currentBest + m$. Here, $currentBest$ is the smallest objective value among all the centers that were computed before.

\subparagraph{Choosing centers.}
Apart from input triangulations, we also generate candidate centers by greedily flipping edges in input triangulations to minimize total intersections with the entire point set until a local optimum is reached.

We then record all the intermediate triangulations at the end of each parallel flip, which is used to transform an input triangulation into one of those centers. To avoid trying all centers, we sort them in increasing order of total number of intersections with all the input triangulations and only compute the heuristic parallel flip distance for a fraction of them.

\section{Combined Strategies}
\subparagraph{Fixed center.}

A promising combined strategy involves using the heuristic to select a candidate center triangulation, denoted as $T_{heur}$, and using the exact SAT solver to compute the minimal parallel flip distances from all input triangulations to $T_{heur}$, one by one. 

Although this approach remains heuristic, it was effective on the \textit{rirs} instances, but it also indirectly helped significantly with large non-\textit{rirs} instances. Finding a slightly better solution allows the exact solver to handle cases that were previously out of reach.

\subparagraph{Solving a subproblem.}

A natural extension to computing pairwise distances is to compute exact distances for larger subsets. This yielded negligible improvements to the lower bound, likely due to the triangle inequality holding strong even for pairwise constraints.

Similar reasoning proved highly effective for the \textit{woc} (especially \textit{tsplib}) instances, observing that the distances to the optimal center were not uniformly distributed. By identifying and excluding outlier triangulations, we could exactly solve for the center of the remaining subset. This produced a robust heuristic center for the fixed center heuristic.

\subparagraph{Solution extension}
For the instance \verb{random_instance_552_320_20{ we also used the following strategy. Start from an optimal solution for the last $10$ triangulations in the input. Then, add one by one the other triangulations. When a triangulation is added, initialize its distance in such a way that the pairwise distances are respected.
Then increase by $1$ the value of the smallest distance in the distance sequence until a satisfiable sequence is found.\\
This method was only used for this instance because we came up with it when there was very little time left in the competition. Moreover, since we did not have to add all the triangulations, we computed an exact solution for the center computed for the last $15$ triangulations by the method we just described. This lead to an improved solution, with parallel flip distance $126$, while the previous best solution we had computed had value $131$.

\section{Future work \& Conclusion}

We have presented a combined approach for computing the central triangulation, combining an exact SAT-based formulation for smaller instances with heuristics for larger instances. Our exact solver effectively handles geometric constraints, while the heuristic strategies allow us to navigate the massive search spaces of larger instances.

\paragraph{Comparison.}
Due to the strict deadline of the competition, we were unable to explore all potential heuristics or fine-tune all proposed ideas. Consequently, several promising approaches remain open for future work. While our method performed well on most instances, the largest \textit{rirs} instances remain a formidable challenge.\\
The competition was won by the team \textit{Shadoks} \cite{shadoks2026}. For smaller and medium-sized instances ($n \le 320$), their approach shares significant similarities with ours, relying on a SAT formulation to find the exact central triangulation. While they employed slightly different clause pruning techniques, the primary differentiator in this regime appears to be the  computational resources they had to solve these hard instances to optimality.\\
However, for the larger instances, specifically the \textit{rirs} instances where $n \geq 500$, they diverged completely from the exact solver paradigm, adopting a heuristic strategy based on Simulated Annealing. In this phase, the search space of the central triangulation is explored by iteratively performing random edge flips. A critical component of their success was the efficiency of the evaluation step: instead of recomputing the flip distances from scratch (which would require a computationally expensive BFS traversal), they utilized an incremental update scheme. Since a single edge flip alters the distance to any target triangulation $T_i$ by exactly $\pm 1$, they could update the total cost in $\mathcal{O}(m)$ time, where $m$ is the number of input triangulations. This rapid evaluation allowed them to explore a vast number of states within the time limit, far exceeding the throughput of a standard exact solver.

\paragraph{Iterative Path Optimization.}
A promising improvement to the upper bounds generated by our heuristics on \textit{rirs} data is to apply a local search on the computed paths to the center. By isolating small segments of the transformation history (e.g., $T^{(i)} \to \dots \to T^{(i+k)}$), we can check if a "shortcut" exists. Leveraging the linear-time 2-SAT formulation from Section~\ref{sec:2sat}, we can efficiently test if a transition of length $k$ can be compressed into $k-1$ steps (specifically for $k=3$ reducing to 2). This iterative tightening could reduce the total distance cost without the overhead of the full SAT solver.

\paragraph{Theoretical Conjecture.}
Beyond the immediate scope of the competition, we intend to focus on resolving the open conjecture concerning the optimal distribution of distances to the center. Our empirical results strongly suggest its validity. A promising direction for a formal proof involves modeling the flip distance behavior on  the Law of Large Numbers and the known bounds of the flip graph's diameter. We hypothesize that by formulating the likelihood of a distribution being to the center being optimal under this distance model, the resulting maximizer will algebraically align with the formula proposed in the conjecture description.

\newpage

\bibliographystyle{plainurl}
\bibliography{bibl_conf}

\end{document}